\documentclass[aps,pra,showpacs,groupedaddress]{revtex4}

\usepackage{graphicx}
\usepackage{dcolumn}
\usepackage{bm}

\begin{document}


\title{Degree of entanglement in a quantum measurement process}
\author{Matthias Jakob}
\author{Yonatan Abranyos}
\author{J\'anos A.\ Bergou}
%
\affiliation{Department of Physics and Astronomy,
Hunter College at the City University of New York, \\
695 Park Avenue New York, NY 10021}
\date{\today}

\begin{abstract}
We suggest a quantum measurement model in an ion trap which specifies
the probability distribution of two, distinct internal ground states
of a trapped four-level ion.
The external degrees of motion of the four-level ion constitute the meter
which, in turn, is coupled to the environment by engineered reservoirs.
In a previous publication, a similar measurement model
was employed to test decoherence effects on quantum nonlocality
in phase space on the basis of coincidence measurements of the entangled
system-meter scheme. Here, we study the effects of decoherence on
the entanglement of formation characterized by the concurrence.
The concurrence of the system enables to find the maximum possible
violation of the Bell inequality.
Surprisingly, this model gives illustrative insights
into the question to what extend the Bell inequality
can be considered as a measure of entanglement.
\end{abstract}

\maketitle
\section{Introduction}\label{Seq.1}
Entanglement constitutes the single most characteristic property
that makes quantum mechanics distinct from any classical theory
\cite{Schroedinger}. It has been in the center of interest since the
early days of quantum theory, mainly due to its relation
to nonlocality \cite{EPR}. It was the discovery of the Bell
inequalities \cite{Bell} that opened the possibility to test nonlocality
in laboratory experiments \cite{Freedman,Aspect}. Surprisingly,
entanglement has found its application in a newly emerging area of
research. It has been recognized that entanglement forms a fundamental
resource for quantum information processing (QIP) \cite{ekert,bennett}.
For application's purposes, it became essential to quantify entanglement
and, accordingly, a number of useful measures of the degree
of entanglement have been introduced \cite{wootters}.
One of them, the {\textit{entanglement of formation}} \cite{wootters1},
is the subject of this paper. The entanglement of formation quantifies
the resources needed to create a given entangled state.

In this paper we study the dynamics of the entanglement of formation
in a quantum measurement model which is a slightly
modified version of our model, proposed previously \cite{jak1}.
We are interested in how the entanglement of formation
is created and lost during the course of the measurement.
Environmentally induced decoherence prevents us from
observing quantum effects in the macroscopic world
\cite{Giulini,zurek1,zurek2,zurek2a,zurek3,walls1,walls2}.
However, we have shown that it is possible to greatly reduce
the decoherence rate by assuming the
environment to be a squeezed reservoir \cite{jak1}.
This enables us to observe violations of Bell-type inequalities
in phase space even when the meter is already in the macroscopic domain.
The entanglement of formation and, in particular, the closely related
concurrence allows us to find the full time-dependence
of the maximum possible violation of Bell-type inequalities.
We will show how the dynamics of the concurrence is affected by
coupling to a squeezed reservoir and present an analytical study of the
dynamics  of the entanglement of formation of decoherent, non-orthogonal
and mixed qubits exposed to an open environment.

We also address the question as to what extent violations of the Bell
inequalities represent a good measure of the degree of entanglement. We
find that not only the degree of mixture of bipartite qubits
plays a role but also the degree of overlap between them. In particular,
the meter generates a non-orthogonal and mixed qubit which can
be rewritten as a superposition of orthogonal qubit states.
Surprisingly, in this context, violations of the Bell inequalities
can be stronger in a system which, compared to other systems, has a larger
degree of mixture but the same amount of entanglement of formation.
We explain this apparently contradictory behavior on the basis
of the measurement model.

The outline of the paper is as follows:
First, in Sec.\ \ref{Sec.2}, we present the measurement model which
consists of a four-level atom in a Paul trap coupled to engineered
reservoirs. Then, in Sec.\ \ref{Sec.3}, we derive the concurrence and
the entanglement of formation and discuss the time evolution of these
quantities for different squeezing parameters
of the environment. We also analyze the role of the Bell inequality as
a measure of entanglement and explain some counterintuitive findings on
the basis of indistinguishability between qubit states.
In Sec.\ \ref{Sec.4} we conclude with some discussion.

\section{Measurement model in a Paul trap}\label{Sec.2}

\begin{figure}[ht,floatfix]
\includegraphics[width=8.6cm]{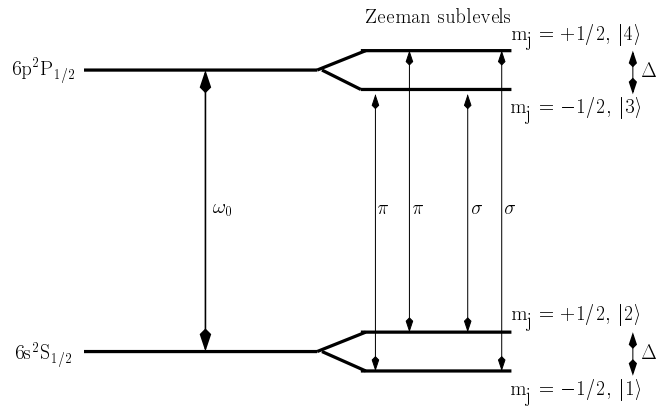}
\caption{\label{Fig. 1}Internal structure of the four-level atom with relevant
polarization dependent
transitions. $\pi$-transitions are responsible for
$z$-polarized light and
$\sigma$-transitions for $x$- or $y$-polarized light.
$m_{j}$ is the magnetic quantum number of the Zeeman sublevels.
This four-level atom has the same structure as the 194-nm transition
from the ground $6{\text{s}}^2{\text{S}}_{1/2}$ state to the
$6{\text{p}}^2{\text{P}}_{1/2}$ level of a trapped
$^{198}\text{Hg}^{+}$-ion in a magnetic field. $\omega_{0}$ is the
transition 
frequency of the $\pi$-transition and $\omega_{0}\pm\Delta$ is the
same for the $\sigma$-transitions, where  $\Delta$ stands for the
detuning between the Zeeman sublevels.}
\end{figure}
\begin{figure}[ht,floatfix]
\includegraphics[width=12cm]{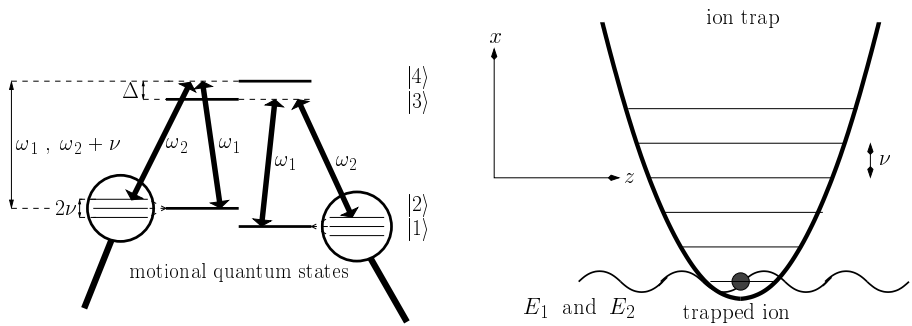}
\caption{\label{Fig. 2}Setup for the quantum measurement 
on the atom. Both Raman lasers, $E_{1,2}$, 
are $x$-polarized ($\sigma$-transitions) and
propagate in the $z$-direction, affecting the quantum motion of the
trapped ion only in that direction. The trap frequency is given by
$\nu$.}
\end{figure}

The measurement model consists of a four-level atom harmonically bound in
a three-dimensional trap where it oscillates along the three
principal axes
with frequencies $\nu_{1}=\nu_{2}=\nu_{3}=\nu$. The level structure of
the four-level atoms with its relevant polarization-sensitive dipole
transitions is shown in Fig.\ \ref{Fig. 1} \cite{Eichmann}. 
We determine the population distribution of the internal atomic
ground-states which, in turn, form the system to be measured.
The external or motional degrees of freedom establish the meter
and its coupling to the environment is accomplished by engineered
reservoirs \cite{myatt1,myatt2}. The trapped four-level atom is driven
in a Raman configuration with two classical,
$\sigma$-polarized laser fields of frequencies $\omega_{1}$
and $\omega_{2}=\omega_{1}+\nu=\omega_{0}$.
The laser frequencies are off-resonant with respect to the electronic
$\sigma$-transitions $|1\rangle\leftrightarrow|4\rangle$
and $|2\rangle\leftrightarrow|3\rangle$ by a detuning of
$\Delta$ ($\Delta\ll\omega_{1},\omega_{2}$), see Fig.\ \ref{Fig. 2}.
A similar ``system-meter'' interaction was investigated by Wallentowitz
and Vogel \cite{wallentowitz}, in order to realize the quantum-mechanical
counterpart of nonlinear optical phenomena in the motional (mechanical)
degrees of freedom. In contrast to our four-level system, however, theirs
consists of a two-level atom where polarization dependent features do
not play a role.

With an appropriate geometry of the lasers the motion can only be affected
in $z$-direction assuming that both laser fields are traveling in
$z$-direction (see Fig. \ref{Fig. 2}). 
Since the lasers are strongly detuned from
the atomic transition frequencies the atom stays in its ground states
during the interaction.
Further, if we assume the resolved sideband limit, we are able to influence
the motional quantum state of the atom in a controlled manner.
In particular, the system-meter interaction Hamiltonian, $H_{\text{SM}}$,
for the given geometry and frequencies of the lasers in the vibrational
rotating-wave approximation is given as \cite{wallentowitz}
\begin{equation}
H_{\text{SM}}=\frac{1}{2}\tilde{\sigma}_{z}
(i\hbar\Omega\eta \hat{a}-i\hbar\Omega^{\ast}\hat{a}^{\dagger}).
\label{Eq2.1}
\end{equation}
Here,
\begin{eqnarray}
\Omega &=& \frac{\Omega_{1}\Omega_{2}^{\ast}}{2\Delta},
\label{Eq2.2} \\
\Delta &=& \omega_{23}-\omega_{0},
\label{Eq2.2a} \\
\Omega_{i} &=& \frac{2d E_{i}}{\hbar},
\label{Eq2.3} \\
\tilde{\sigma}_{z} &=&
|2\rangle\langle 2|-|1\rangle\langle 1|+|4\rangle\langle 4|
-|3\rangle\langle 3|,
\label{Eq2.4}
\end{eqnarray}
where $\eta$ is the Lamb-Dicke parameter, $d$ is the dipole moment which
we assume
to be the same for all possible dipole transitions in the four-level atom,
and $E_{i}$ are the electric field amplitudes of the applied lasers.
We have assumed a small Lamb-Dicke parameter, $\eta \ll 1$, in the
interaction Hamiltonian which allows us to neglect nonlinear terms in the
motional operators $\hat{a}$ and $\hat{a}^{\dagger}$. The phase of
$\Omega = \Omega e^{i\phi}$ can be adjusted by the phase difference
of the two lasers. From now on, we assume $\phi=\pi$ which leads
to the following interaction Hamiltonian
\begin{equation}
H_{\text{SM}}=\frac{i}{2}\hbar\tilde{\sigma}_{z}
(\eta|\Omega|\hat{a}^{\dagger}-\eta|\Omega|\hat{a}).
\label{Eq2.5}
\end{equation}
This type of system-meter Hamiltonian is of the same structure as the
interaction Hamiltonian between a four-level atom and a cavity field
which we recently employed to study decoherence effects on the visibility
of interference fringes \cite{abra1} and on nonlocality in phase space
\cite{jak1}.

The replacement of the optical meter by the mechanical one in the
system-meter interaction has considerable advantages in the realization
of the measurement model. In particular, the experimental progress in
reservoir engineering
for trapped ions \cite{myatt1,myatt2} makes this measurement model
a feasible testing ground of decoherence effects in engineered reservoirs.
Further, the interaction Hamiltonian (\ref{Eq2.1}) can be easily modified
to include nonlinear terms in the motional operators
$\hat{a}$ and $\hat{a}^{\dagger}$ by appropriate settings of the frequency
and geometry of the applied lasers.
The nonlinear interaction of parametric type together with specific
reservoirs, such as dissipative two-phonon processes,
makes it possible to generate nonclassical, macroscopic motional states
in dissipative environments (Gilles and Knight \cite{gilles}).

Here, however, we do not consider nonlinear effects in the system-meter
interaction.
The environment, as in Ref.\ \cite{jak1}, is taken to be a squeezed
reservoir which can be engineered according to Refs.\ \cite{myatt1,myatt2}.
The master equation for the system-meter density operator in the Markov
approximation is given as
\begin{eqnarray}
\frac{\partial}{\partial t}\rho&=&\frac{1}{i\hbar}[H_{\text{SM}},\rho]
+\frac{\gamma}{2}\Bigr\{(N+1)[2b\rho b^{\dagger}-b^{\dagger}b\rho
-\rho b^{\dagger}b]
+N[2b^{\dagger}\rho b-bb^{\dagger}\rho-\rho bb^{\dagger}]
\nonumber \\
&&
+ M[2b^{\dagger}\rho b^{\dagger}-b^{\dagger}b^{\dagger}\rho
-\rho b^{\dagger}b^{\dagger}]+ M^*[2b\rho b-bb\rho-\rho bb]\Bigl\},
\label{Eq2.6}
\end{eqnarray}
where $N$ is the number of photons and $M=-|M|e^{2i\theta}$
is the squeezing parameter which characterizes the degree
of phase-dependent correlations, with the squeezing phase $\theta$,
in the squeezed reservoir.
The master equation can be analytically solved with a characteristic
function approach, leading to the following time evolution of the
system-meter density operator \cite{jak1}
\begin{equation}
\rho(t)=\sum_{n,m=1}^{2}\rho_{nm}e^{\Gamma^{\text{sq}}_{nm}(t)}
|n\rangle\langle m|\otimes\frac{|\tilde{\alpha}_{n}(t),\varepsilon\rangle
\langle\tilde{\alpha}_{m}(t),\varepsilon|}
{\langle\tilde{\alpha}_{m}(t),\varepsilon|\tilde{\alpha}_{n}(t) .
\varepsilon\rangle}, \label{Eq2.7}
\end{equation}
The summation is over the internal atomic ground states, $|1\rangle$ and
$|2\rangle$, and $\rho_{nm}$ are the initial atomic density matrix
elements which we assume to be given as $\rho_{nm}=1/2$ for all $n,m$.
The amplitudes of the ensuing squeezed coherent states,
$|\alpha,\varepsilon\rangle=\exp({\frac{1}{2}\varepsilon^{\ast}\hat{a}^2-
\frac{1}{2}\varepsilon\hat{a}^{\dagger 2})}|\alpha\rangle$,
are given by
\begin{eqnarray}
\tilde{\alpha}_{n}(t)&=&\frac{|\Omega|\eta (-1)^{n}}{\gamma}
\left[\cosh(r)-\sinh(r)e^{i2\theta}
\right]\left(1-e^{-\frac{\gamma t}{2}}\right).
\label{Eq2.8}
\end{eqnarray}
Here $\varepsilon=re^{2i\theta}$ with the squeezing parameter
$r$ and the squeezing phase $\theta$ defined in the usual way.
These states form the pointer states of the meter.

The squeezed coherent states $|\tilde{\alpha}_{n}(t),\varepsilon\rangle$,
$n=1,2$, follow from first displacing the vacuum by the amplitude
$[(-1)^{n}|\Omega|\eta/\gamma](1-e^{-\frac{\gamma t}{2}})$
and then squeezing the resulting coherent state.
The exponent $\Gamma_{nm}^{\text{sq}}(t)$ is responsible for the
decoherence and is given as
\begin{eqnarray}
\Gamma^{\text{sq}}_{nm}(t)&=&\Bigl\{\frac{[(-1)^{n}-(-1)^{m}]^2
(|\Omega|\eta)^2}{\gamma^2}
\left\{1+2\left[\sinh^2(r)-\cosh(r)\sinh(r)\cos2(\theta)\right]\right\}
\Bigl(1-\frac{\gamma t}{2}-e^{-\frac{\gamma t}{2}}\Bigr)\Bigr\},
\label{Eq2.9}
\end{eqnarray}
In Ref.\ \cite{jak1} we demonstrated how to reduce the decoherence
rate of the measurement apparatus with the help of the squeezed reservoir
by adjusting the squeezing phase to $\theta=0$ and increasing the
squeezing parameter $r$. 
\begin{figure}[h,t]
\includegraphics[width=8.6cm]{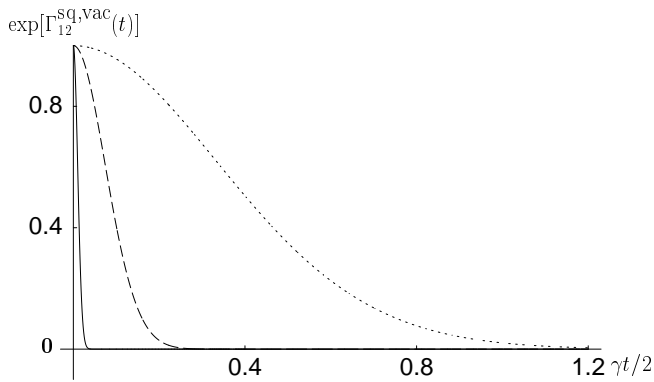}
\caption{\label{Fig. 3}Comparison of the decoherence,
$\exp[\Gamma_{12}^{\text{sq,vac}}(t)]$,
in a squeezed reservoir (sq) to that in an ordinary vacuum (vac)
(full line)
vs.\ the dimensionless time, $\gamma t/2$.
The squeezing parameters are given
by $r=2$ and $\theta=0$ (dashed line) and $r=3.5$ and
$\theta=0$ (dotted line).
The amplitudes of the corresponding squeezed states (see text) are
given by $\pm\alpha_{0}=100$, respectively.}
\end{figure}
This is displayed in Fig.\ \ref{Fig. 3}
where the amplitude of the ensuing squeezed coherent state
$|\tilde{\alpha},\varepsilon\rangle$ is given as
$\alpha_{0}=|\Omega|\eta/\gamma=100$.
The amplitude can be engineered by suitable settings of 
the laser strengths and the engineered reservoir
damping constant $\gamma$ keeping in mind that the constraints $\eta\ll
1$ and $\Omega_{i}/2\Delta\ll 1$ (with $i=1,2$) must be satisfied.
In spite of these restrictions, it is possible to achieve large
amplitudes, $\Omega\eta/\gamma \gg 1$.
As a result of this highly reduced decoherence rate, we could predict
the existence of distinctive quantum features of the meter even
in a macroscopic domain. Moreover, we studied nonlocal properties of the
coupled
system-meter scheme and demonstrated violations of Bell-type inequalities
\cite{Bell,Bell1} in the Clauser-Horne-Shimony-Holt form \cite{Clauser}
of phase space observables \cite{wodk1a,wodk2} even when the
meter reached a macroscopic state \cite{jak1}.

\section{Entanglement of formation}\label{Sec.3}

Underlying nonlocality is the concept of entanglement. It is
responsible for the fact that a composite system possesses properties
which can not be understood by considering
the parts of the system separately. In other words, there is no element of
``reality'' in the parts considered by their own but only a created
reality which depends on what is measured in the other part.
This uniquely quantum concept has proved to be a fundamental
resource for quantum information processing and the quantification of
entanglement is essential to assess the full performance of an
information theory based on quantum mechanics \cite{ekert,bennett}.
There are several good measures of the degree of entanglement for both
pure and mixed quantum states \cite{wootters}.
Perhaps the most seminal one is the {\textit{entanglement of formation}}
which quantifies
the resources needed to {\textit{create}} an entangled state
\cite{wootters1}. Entanglement of formation is a measurable quantity, at
least for a pair of qubits which is the case we are dealing with
here. The underlying quantity is called {\textit{concurrence}}
\cite{wootters1}. For pure states, concurrence is strongly connected
with two-particle visibility \cite{jaeger,abouraddy} which is
a property that cannot exist separately in the parts of a bipartite system.
The expression relating the concurrence to the density operator,
$\rho$, of a mixed state is given as
\begin{equation}
C(\rho)={\text{max}}\{0,\lambda_{1}-\lambda_{2}-\lambda_{3}-\lambda_{4}\},
\label{Eq2.10}
\end{equation}
where the $\lambda_{i}$'s are the square roots of the eigenvalues of
$\rho\tilde{\rho}$ in descending order. Here $\tilde{\rho}$ results from
applying the spin-flip operation to $\rho^{\ast}$,
\begin{equation}
\tilde{\rho}=(\sigma_{y}\otimes\sigma_{y})\rho^{\ast}
(\sigma_{y}\otimes\sigma_{y}) .
\label{Eq2.11}
\end{equation}
Here $\sigma_{y}$ is the Pauli spin operator in the standard basis and
$\rho^{\ast}$ is the complex conjugate of $\rho$.
The entanglement of formation, $E_{f}(\rho)$, of the state $\rho$
is connected with the concurrence, $C(\rho)$, via the formula
\begin{eqnarray}
E_{f}(\rho)&=& h(\frac{1+\sqrt{1-C^2(\rho)}}{2}), \label{Eq2.12} \\
h(x)&=&-x\log_{2}x-(1-x)\log_{2}(1-x). \label{Eq2.13}
\end{eqnarray}
There is another remarkable property of the concurrence which is directly
related to nonlocality. One can show that the maximum possible violation
of the Bell inequality \cite{Bell,Bell1} in the Clauser-Horne-Shimony-Holt
form \cite{Clauser}
\begin{equation}
{\cal{B}}(\rho)=|E(c,d)+E(c',d)+E(c,d')-E(c',d')|\leq 2,
\label{Eq2.14}
\end{equation}
for a state $\rho$ is given as \cite{abouraddy,gisin,horodecki,ghosh}
\begin{equation}
{\cal{B}}_{\text{max}}(\rho)=2\sqrt{M(\rho)}.
\label{Eq2.15}
\end{equation}
Here $c,c'$ are two dichotomous variables of the first system
and $d,d'$ are two of the second, and $E(c,d)$ is the expectation value
of the correlation of $c$ and $d$, and so on for the other expectation
values. The quantity $M(\rho)$ is the sum of the two larger eigenvalues
of $T_{p}T_{p}^{\dagger}$, where $T_{p}$ is the $3\times3$ matrix
with the $(m,n)$ element given by
\begin{equation}
t_{mn}=\text{tr}(\rho\sigma_{n}\otimes\sigma_{m}), \label{Eq2.15a}
\end{equation}
and the $\sigma_{i}$'s are the Pauli matrices.

Based on the previous considerations, a number of interesting questions
arise with respect to our measurement model concerning the generation
of the entanglement of formation and its dynamical properties.
Among them is the control of the dynamical properties of the entanglement
of formation and the possibility of observing its quantum features in the
macroscopic domain of the meter in a squeezed reservoir environment.
Another one is the determination of the maximum possible violation of the
Bell inequality (\ref{Eq2.14}) and the time when it is achieved in the
measurement.

To answer these questions it is essential to recognize that the potentially
entangled system-meter state consists of {\textit{nonorthogonal}}
``qubits''. In particular, the squeezed coherent states
$|\tilde{\alpha}_{n}(t),\varepsilon\rangle$ of the meter are not completely
orthogonal during the course of the measurement process. It is, however,
possible to define concurrence and, consequently, entanglement of formation
for nonorthogonal bipartite systems by introducing an orthonormal basis
in the two subsystems of the bipartite state \cite{wang,hirota}.
In the measurement model under consideration the orthonormal, time-dependent
basis for the meter is formed by
\begin{eqnarray}
|\tilde{0}(t)\rangle &=& |-\tilde{\alpha}(t),\epsilon\rangle, \label{Eq2.16}
\\
|\tilde{1}(t)\rangle &=&\frac{ |\tilde{\alpha}(t),\epsilon\rangle
-P(t)|-\tilde{\alpha}(t),\epsilon \rangle }
{\sqrt{1-|P(t)|^2}},
\label{Eq2.17} \\
\tilde{\alpha}(t) &=& \alpha_{0}\left[\cosh(r)-\sinh(r)\right]
\left(1-e^{-\frac{\gamma t}{2}}\right), \label{Eq2.18} \\
P(t) &=& |P(t)| = \langle-\tilde{\alpha}(t),\epsilon|\tilde{\alpha}(t),
\epsilon\rangle.
\label{Eq2.18a}
\end{eqnarray}
Here we set the squeezing phase $\theta=0$ in order to maximize the effect
of the 
squeezed reservoir on the decoherence rate (see Fig.\ \ref{Fig. 3}).
The orthonormal basis states for the system, of course, are given by the
two ground states of the four-level atom, $|1\rangle$ and $|2\rangle$.
We can now construct the spin-flip operators for the meter,
$\sigma_{y}^{\text{M}}(t)$, and for the system,
$\sigma_{y}^{\text{S}}(t)$, as
\begin{eqnarray}
\sigma_{y}^{\text{M}}(t) &=& i \left\{|\tilde{1}(t)\rangle\langle
\tilde{0}(t)| - |\tilde{0}(t)\rangle\langle\tilde{1}(t)| \right\},
\label{Eq2.19} \\
\sigma_{y}^{\text{S}} &=& i \left\{|2\rangle\langle 1|-|1\rangle\langle
2| \right\}.
\label{Eq2.20}
\end{eqnarray}
With these results it is straightforward to calculate the concurrence
(\ref{Eq2.10}), from Eqs.\ (\ref{Eq2.11}) and (\ref{Eq2.7}) of the
system-meter state, in spite of its complicated dynamical properties.
Its analytical expression takes a rather simple form,
\begin{equation}
C(\rho(t))=\exp[\Gamma_{12}^{\text{sq}}(t)]\frac{\sqrt{1-P^{2}(t)}}{P(t)}.
\label{Eq2.21}
\end{equation}
Surprisingly, it is possible, at least in principle, to determine the
concurrence of the system-meter state by directly measuring the
(time-dependent) operator
$\sigma_{y}^{\text{M}}(t)\otimes\sigma_{y}^{\text{S}}$
\begin{equation}
C(\rho(t)) = -\langle\rho(t)\sigma_{y}^{\text{M}}(t) \otimes
\sigma_{y}^{\text{S}} \rangle.
\label{Eq2.22}
\end{equation}
When we recall that our measurement model contains a rather involved
decoherence mechanism for the system-meter and, therefore, its state
rapidly approaches a mixture, this is an amazing result that gives the
concurrence (at least in this model) an operational meaning.
Analogously, we can calculate the maximum violation
of the Bell inequality, Eq.\ (\ref{Eq2.15}), yielding
\begin{equation}
{\cal{B}}_{\text{max}}(\rho(t))=2\sqrt{1+C^2(\rho(t))+
\exp[2\Gamma_{12}^{\text{sq}}(t)]-P^{2}(t)}.
\label{Eq2.22a}
\end{equation}

\begin{figure}[h,t]
\includegraphics[width=14cm]{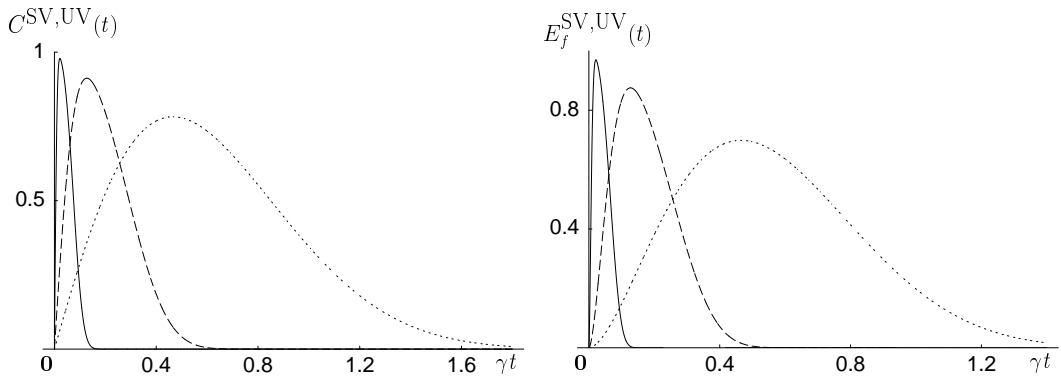}
\caption{\label{Fig. 4}Time dependence of the concurrence,
$C^{{\text{SV,UV}}}(\rho,t)$, (left figure)
and entanglement of formation, $E_{f}^{{\text{SV,UV}}}(\rho,t)$,
(right figure) for different squeezing parameters.
The notation as well as the parameters
are the same as in Fig.\ \ref{Fig. 3}. SV denotes squeezed
vacuum and UV denotes ordinary vacuum.}
\end{figure}
\begin{figure}[h,t]
\includegraphics[width=14cm]{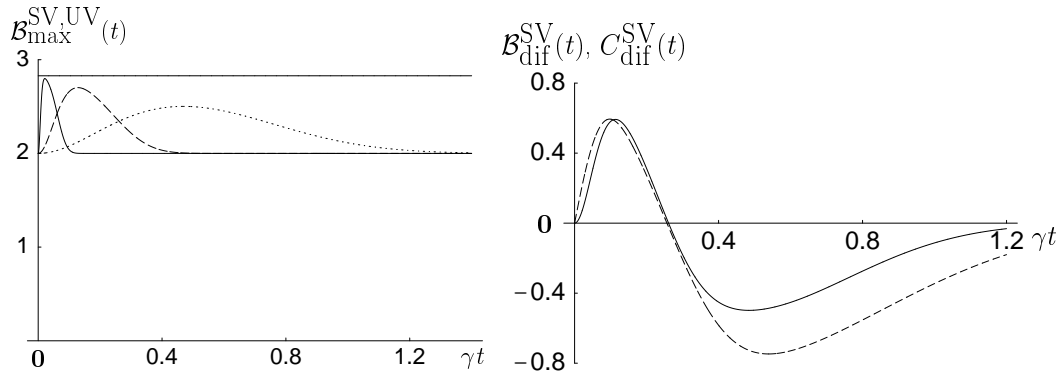}
\caption{\label{Fig. 5}Time-dependence of the maximum 
possible violation of the
Bell-inequality for different squeezing parameters (left figure).
The notation as well as the parameters
are the same as in Fig.\ \ref{Fig. 3}.
The upper straight line represents the maximum possible
violation of the Bell inequality by quantum mechanics.
The right figure displays the time dependence of the difference
between the maximum possible violations of the Bell inequality (full
line) and between the concurrences (dashed line) in two different
squeezed 
reservoirs with squeezing parameters $r_{1}=2$, $r_{2}=3.5$ and
$\theta_{1}=\theta_{2}=0$, respectively.}
\end{figure}

Based on Eq.\ (\ref{Eq2.21}), we have plotted the concurrence as well
as the entanglement of formation, Eqs.\ (\ref{Eq2.12}) and
(\ref{Eq2.13}), for different squeezing parameters in Fig.\ \ref{Fig. 4}.
With increasing squeezing parameter, the entanglement of formation
approaches its maximum at a much later time than in a regular vacuum.
A squeezed environment which monitors the meter is capable to maintain
nonclassical properties of the system-meter, i.e.\ the correlations
between them, over a much longer period of time.
The system-meter state has already entered its macroscopic domain when
the entanglement of formation reaches its maximum. This can be
seen in Fig.\ \ref{Fig. 4} for a squeezed reservoir with a squeezing
parameter of $r=3.5$ and for the corresponding squeezed coherent states
of amplitude $\alpha_{0}=100$, for example. In addition, we also see that
the entanglement of formation does not reach its maximum possible value
of $1$, irrespective of the type of reservoir which monitors the
system-meter. This, however, is not surprising for a mixed state
which contains partly nonorthogonal qubits.
The maximum possible value of the entanglement of formation of the
system-meter is additionally reduced in a squeezed environment, since
there is a competition between the positive effect of the highly reduced
decoherence rate (which maintains the purity of the system-meter on a
greatly enhanced time-scale) and the negative effect of the larger overlap
(i.e. the intrinsic indistinguishability of the meter) between the
nonorthogonal meter states. In Fig.\ \ref{Fig. 5} we display the time
evolution of the maximum possible violation of the Bell inequality, Eq.\
(\ref{Eq2.22a}), in the measurement model.
Again, we can observe violations of the Bell inequality on a
greatly enhanced time scale in a squeezed reservoir with
increasing squeezing parameter. We have also found this result in Ref.\
\cite{jak1} based on a phase-space equivalent of the Bell-inequality
\cite{wodk1a,wodk2}.
However, the approach with the concurrence of the system-meter has a
number of advantages. First, we are able to observe the full
time-dependence of
the ``formation'' of entanglement during the course of the measurement.
Second, we can display the maximum possible
violation of the Bell inequality at every time step. In general, it
is hard to find this quantity on the basis of the inequality, Eq.\
(\ref{Eq2.14}), 
which depends on four parameters. Beyond it, the phase-space approach of
Banaszek and W\'odkiewicz \cite{wodk1a,wodk2} can not approach the maximum
possible value of the violation of the Bell inequality because of the
smoothing effect of the Wigner function.

The right insert of Fig.\ \ref{Fig. 5} displays the time dependence of
the difference between the maximum possible violations of the Bell
inequality, Eq.\ (\ref{Eq2.22a}), and between the concurrences, Eq.\
(\ref{Eq2.21}), in two different squeezed reservoir environments with
squeezing parameters $r_{1}=2$, $r_{2}=3.5$ and $\theta_{1}=\theta_{2}=0$. 
From the figure one can get some insight as to what extent a violation
of the Bell inequality tells us something about the nature of entanglement
\cite{ghosh,munro}. Munro {\it et al.} \cite{munro} suggested that the
more mixed a system is made the more entanglement (or concurrence) is
generally required to violate the Bell inequality to the same degree.
This, however, is not generally true as pointed out by Ghosh {\it et al.}
\cite{ghosh} but they could not find a simple explanation.
Based on the right part of Fig.\ \ref{Fig. 5} we can answer this
question as follows. At $\gamma t\approx 0.3$ 
in Fig.\ \ref{Fig. 5} the difference of the concurrences 
between the two systems
$C_{\text{dif}}^{\text{SV}}(t)=C_{1}^{\text{SV}}(t)-C_{2}^{\text{SV}}(t)$
is approximately zero while the difference of the maximum possible
violations of the Bell inequality,
${\cal{B}}_{\text{dif}}^{\text{SV}}(t) =
{\cal{B}}_{\text{max},1}^{\text{SV}}(t) -
{\cal{B}}_{\text{max},2}^{\text{SV}}(t)$, is slightly larger than zero.
In contrast, the degree of mixedness of system $1$, corresponding to a
squeezed environment with
parameters $r_{1}=2$ and $\theta_{1}=0$, is obviously larger
than that of system $2$ (with squeezing parameters
$r_{2}=3.5$ and $\theta_{2}=0$) as a consequence of the
advanced decoherence in system $1$ (see Fig. \ref{Fig. 3}).
Thus, this system contradicts the statement by Munro {\it et al.}
\cite{munro}. We get a larger violation of the Bell
inequality in a system having the same degree of entanglement
but, at the same time, more mixedness than the other system.
We suggest the following explanation for this apparently peculiar
behavior. It is the overlap between the nonorthogonal meter states that
reduces the maximum possible violation, Eq.\ (\ref{Eq2.22a}).
In order to violate the Bell inequality it is necessary for the
components of the state of the composite system to be distinguishable.
When the degree of distinguishability gets smaller the amount of the
maximum possible violation of the Bell inequality will also be reduced.
This explains why the difference of the
maximum possible violations of the Bell inequality in Fig.\ \ref{Fig. 5}
is slightly larger than zero in spite of the fact that system $1$
is more mixed but has the same amount of entanglement as system $2$.
Obviously, the overlap of the meter states of system $2$ at this
particular time is much larger than that of system $1$.
This observation can give a novel direction to investigations of the
problem as to what 
extent the Bell inequality is a measure of entanglement and connects it
not only with mixedness but also with the degree of
indistinguishability of nonorthogonal qubits.

\section{Summary}\label{Sec.4}

In summary, we have investigated dynamical properties of the entanglement
of formation in a measurement model. We have also demonstrated the ability
to influence the time evolution of the entanglement of formation by a
squeezed reservoir and found a way to maintain this nonclassical property
in a macroscopic domain of the meter, in spite of it being monitored by the
environment. Furthermore, this model gives some insight into dynamical
properties of the entanglement of decoherent and nonorthogonal entangled
qubits which is of central interest in quantum information theory
\cite{nielsen}.

Finally, we note that it is possible to implement
this measurement model with a trapped ``four-level''
$^{198}\text{Hg}^{+}$-ion
\cite{Eichmann} which is exposed to engineered reservoirs
\cite{myatt1,myatt2}.
In addition, it seems, at least in principle, possible to directly
measure the concurrence with time-dependent Pauli spin-flip operators.
The present measurement model is well suited to study the entanglement
of formation in dissipative environments and helps to clarify some
of the underlying physical principles.
In particular, it gives new insights into the question as to what extent
the Bell inequality is a measure of entanglement and explains
how to get larger amounts of violation in a system with more mixedness
but the same amount of entanglement as a reference system.

\begin{acknowledgments}
We acknowledge discussions with M.\ Hillery, I.\ Nemeth and Y.\ Sun.
This research was supported by a grant from the Office of Naval Research
(Grant No.\ N00014-92J-1233) and by a grant from PSC-CUNY.
\end{acknowledgments}

\begin{thebibliography}{99}
\bibitem{Schroedinger} E.\ Schr\"odinger, Naturwissenschaften {\bf 23}, 807
(1935);
{\bf 23}, 823 (1935); {\bf 23}, 844 (1935).
\bibitem{EPR} A.\ Einstein, B.\ Podolsky, and N.\ Rosen,  Phys.\ Rev.\ 
{\bf 47}, 777 (1935). 
\bibitem{Bell} J.\ S.\ Bell, Physics (Long Island City, NY)
{\bf 1}, 195 (1964).
\bibitem{Freedman}S.\ J.\ Freedman and J.\ F.\ Clauser, \prl {\bf 28}, 938
(1972).
\bibitem{Aspect} See for example: A.\ Aspect, Nature {\bf 398}, 189 (1999);
P.\ Grangier, {\textit{ibid.}} {\bf 409}, 774 (2001), and references
therein.
\bibitem{ekert} A.\ Ekert, \prl {\bf 67}, 661 (1991).
\bibitem{bennett} C.\ H.\  Bennett, G.\ Brassard, C.\ Crepeau,
R.\ Josza, A.\ Peres, and W.\ K.\ Wooters, \prl {\bf 70}, 1895 (1993).
\bibitem{wootters} For a good overview see:
W.\ K.\ Wootters, Quant.\ Inf.\ Comp.\ {\bf 1}, (1) 27 (2001),
and references therein.
\bibitem{wootters1} W.\ K.\ Wootters, \prl {\bf 80}, 2245 (1998).
\bibitem{jak1} M.\ Jakob, Y.\ Abranyos, and J.\ A.\ Bergou,
\pra {\bf 64}, 062102 (2001).
\bibitem{Giulini}
D.\ Giulini, E.\ Joos, C.\ Kiefer, J.\ Kupsch, I.-O.\ Stamatescu, H.-D.\
Zeh,
{\textit{Decoherence and the Appearance of a Classical World in Quantum
Theory}},
(Springer, Berlin, 1996).
\bibitem{zurek1}
W.\ H.\  Zurek, \prd $\bf{24}$, 1516 (1981).
\bibitem{zurek2}
W.\ H.\ Zurek, \prd $\bf{26}$, 1862 (1982).
\bibitem{zurek2a}
W.\ G.\ Unruh and W.\ H.\ Zurek, \prd ${\bf 40}$, 1071 (1989).
\bibitem{zurek3}
W.\ H.\ Zurek, Phys.\ Today {\bf{44}}, 36 (1991), and references therein.
\bibitem{walls1}
D.\ F.\ Walls, M.\ J.\ Collett, and G.\ J.\ Milburn,
\prd $\bf{32}$, 3208 (1985).
\bibitem{walls2}
D.\ F.\ Walls and G.\ J.\ Milburn, {\textit{Quantum Optics}},
(Sprin\-ger, Berlin, 1994).
\bibitem{Eichmann} U.\ Eichmann, J.\ C.\ Bergquist, J.\ J.\ Bollinger,
J.\ M.\ Gilligan, W.\ M.\ Itano, D.\ J.\ Wineland, and M.\ G.\ Raizen,
\prl {\bf 70}, 2359 (1993); W.\ M.\ Itano, J.\ C.\ Bergquist,
J.\ J.\ Bollinger, D.\ J.\ Wineland, U.\ Eichmann, and M.\ G.\ Raizen,
\pra {\bf 57}, 4176 (1998).
\bibitem{myatt1}
C.\ J.\ Myatt, B.\ E.\ King, Q.\ A.\ Turchette,
C.\ A.\ Sackett, D.\ Kielpinski, W.\ M.\ Itano,
C.\ Monroe, and D.\ J.\ Wineland, Nature (London) $\bf{403}$, 269 (2000);
Q.\ A.\ Turchette, C.\ J.\ Myatt, B.\ E.\ King,
C.\ A.\ Sackett, D.\ Kielpinski, W.\ M.\ Itano,
C.\ Monroe, and D.\ J.\ Wineland, \pra {\bf 62}, 053807 (2000).
\bibitem{myatt2}
C.\ J.\ Myatt, B.\ E.\ King, Q.\ A.\ Turchette, C.\ A.\ Sackett,
D.\ Kielpinski, W.\ M.\ Itano, C.\ Monroe, and D.\ J.\ Wineland,
J.\ Mod.\ Opt.\ {\bf{47}}, 2181 (2000).
\bibitem{wallentowitz} S.\ Wallentowitz and W.\ Vogel, \pra ${\bf{55}}$,
4438 (1997).
\bibitem{abra1}
Y.\ Abranyos, M.\ Jakob, and J.\ A.\ Bergou, \pra $\bf{60}$, R2618 (1999).
\bibitem{gilles}
L.\ Gilles, B.\ M.\ Garraway, and P.\ L.\ Knight, \pra {\bf 49}, 2785
(1994).
\bibitem{Bell1} J.\ S.\ Bell,
{\textit{Speakable and Unspeakable in Quantum Mechanics}},
(Cambridge University Press, Cambridge, England, 1987).
\bibitem{Clauser} J.\ F.\ Clauser, M.\ A.\ Horne,
A.\ Shimony, and R.\ A.\ Holt, \prl {\bf 23}, 880 (1969);
J.\ S.\ Bell, in {\textit{Foundations of Quantum Mechanics}},
edited by B.\ d'Espagnat (Academic, New York, 1971);
J.\ F.\ Clauser and A.\ Shimony, Rep.\ Prog.\ Phys.\ {\bf 41}, 1881 (1978),
and references therein.
\bibitem{wodk1a}  K.\ Banaszek and K.\ W\'odkiewicz,
\pra {\bf 58}, 4345 (1998).
\bibitem{wodk2} K.\ Banaszek and K.\ W\'odkiewicz, \prl {\bf 82}, 2009
(1999).
\bibitem{jaeger}G.\ Jaeger, M.\ A.\ Horne, and A.\ Shimony, \pra {\bf 48},
1023 (1993);
G.\ Jaeger, A.\ Shimony, and L.\ Vaidman, \pra {\bf 51}, 54 (1995).
\bibitem{abouraddy} A.\ F.\ Abouraddy, B.\ E.\ A.\ Saleh, A.\ S.\ Sergienko,
and M.\ C.\ Teich, \pra {\bf 64}, 0505101(R), (2001).
\bibitem{gisin} N.\ Gisin, Phys.\ Lett.\ A {\bf 154}, 201 (1991); S.\
Popescu 
and D.\ Rohrlich, {\textit{ibid.}} {\bf 166}, 293 (1992).
\bibitem{horodecki} M.\ Horodecki, P.\ Horodecki, and R.\ Horodecki, Phys.\
Lett.\ A {\bf 200}, 340 (1995).
\bibitem{ghosh} S.\ Ghosh, G.\ Kar, A.\ Sen(De), and U.\ Sen,
\pra {\bf 64}, 044301, (2001).
\bibitem{wang} H.\ Fu, X.\ Wang, and A.\ Solomon,
quant-ph/0105099; X.\ Wang, quant-ph/0102011.
\bibitem{hirota} O.\ Hirota, S.\ J.\ van Enk, K.\ Nakamura, M.\ Sohma,
and K.\ Kato, quant-ph/0101096;
O.\ Hirota and M.\ Sasaki, {\textit{ibid.}}, quant-ph/0101018.
\bibitem{munro} W.\ J.\ Munro, K.\ Nemoto, and A.\ G.\ White,
\jmo {\bf 48}, 1239 (2001).
\bibitem{nielsen} see for example: M.\ A.\ Nielsen and I.\ L.\ Chuang,
{\textit{Quantum Compution and Quantum Information}},
(Cambridge University Press, Cambridge 2000), and references therein.
\end{thebibliography}
\end{document}